\documentclass[conference]{IEEEtran}
\usepackage{amsmath}
\usepackage{amsfonts}
\usepackage{graphicx}
\usepackage{array}
\usepackage{verbatim}

\title{Model Checking in multiplayer games development}

\author{\IEEEauthorblockN{Ruslan Rezin}
\IEEEauthorblockA{
\textit{Innopolis University}\\
Innopolis, Russia \\
r.rezin@innopolis.ru}
\and
\IEEEauthorblockN{Ilya Afanasyev}
\IEEEauthorblockA{ 
\textit{Innopolis University}\\
Innopolis, Russia \\
i.afanasyev@innopolis.ru}
\and
\IEEEauthorblockN{Manuel Mazzara}
\IEEEauthorblockA{
\textit{Innopolis University}\\
Innopolis, Russia \\
m.mazzara@innopolis.ru}
\and
\IEEEauthorblockN{Victor Rivera}
\IEEEauthorblockA{
\textit{Innopolis University}\\
Innopolis, Russia \\
v.rivera@innopolis.ru}
}

\begin{document}

\maketitle
\newpage

\begin{abstract}
Multiplayer computer games play a big role in the ever-growing entertainment industry. Being competitive in this industry means releasing the best possible software, and reliability is a key feature to win the market. Computer games are also actively used to simulate different robotic systems where reliability is even more important, and potentially critical. Traditional software testing approaches can check a subset of all the possible program executions, and they can never guarantee complete absence of errors in the source code. On the other hand, during more than twenty years, Model Checking has demonstrated to be a powerful instrument for formal verification of large hardware and software components. In this paper, we contribute with a novel approach to formally verify computer games. We propose a method of model construction that starts from a computer game description and utilizes Model Checking technique. We apply the method on a case study: the game \textit{Penguin Clash}. Finally, an approach to game model reduction (and its implementation) is introduced in order to address the state explosion problem. 


\end{abstract}

\section{Introduction}


The gap between research in formal verification and its adoption in actual game development is still significant. Despite major conceptual and technological advances, game designers still use the same instruments, and this is not on par with comparable entertainment industries \cite{Almeida2013,Nei12}, or other well-established fields where the discussion spans over tools and methodologies \cite{Gmehlich13,Mazzara2010}. Part of the explanation comes from the scarce usability of modern software verification tools that often still requires help from the experts or good mathematical background from the developers, or possibly both \cite{Khazeev2017}. For testing purposes, Test Driven Development (TDD) \cite{MAD10} is usually used in practice but, since tests cover just a subset of the set of all possible executions of a program, TDD helps only to find bugs, but not to prove their absence. This is where formal verification is necessary. The objective of this paper is bringing formal verification methods into general software development and, in particular, in game development.

The game process usually can be represented by a nondeterministic finite automata (NFA) \cite{RAB59} which evolves each unit of time based on the players' actions or random actions integrated into gameplay. In fact, programmers usually intuitively keep in mind this NFA and implement it implicitly. The more features are added to the game and the bigger the NFA grows, the harder is to test its fidelity. 

From the other hand construction of the model for complex games is not straightforward. The problem comes from the fact that automata for Model Checking is usually represented by a set of first order predicate logic formulas and it is not obvious to write them by hand.

This leads to the idea of developing a tool which can support the construction of a formal representation of the game's NFA (game model), and then automatically verify the correctness of the properties of interest. After that, the game model can automatically be translated into a higher level programming language in order to integrate it into the game under development.

In this paper, we contribute with a novel approach to formally verify multiplayer computer games. We propose a method of model construction that starts from a game description and  utilizes Model Checking \cite{CLA99}. We present implementation of the method, \texttt{Safegame} tool, and demonstrate on a case study: the game \textit{Penguin Clash}. We also show how the tool can be used to create reduced models from the original one so to cope with the state explosion problem.


The paper is structured as follows. Section \ref{sec:rwork} provides state-of-art in formal verification applications to computer game development and robotic systems. Section \ref{sec:gmodel} gives a formal definition of multiplayer games. 
Section \ref{sec:vm} provides the description of the verification method and its application to a case study of the \textit{Penguin Clash} game starting from specification and finishing with verification results of different properties. Finally, an approach of model reduction to address state explosion problem is proposed.

\section{Related work} \label{sec:rwork}

Software verification is a very active research field. However, the area of games has been only partially investigated. In \cite{MAR12}, a case study of verification of role-playing game (RPG) is discussed. The authors use the Algebraic Petri Nets Analyzer (ALPiNA) model checker to verify reachability properties of the model, represented by the Petri Nets. The results of the work are very promising, however the game model has not been formally defined, and the necessary steps for constructing it from the game specification were not explicitly discussed. Moreover, examples of more complex game properties have not been studied.

Computer games storyline description and verification problem was studied in \cite{HOL16}. The objective is to develop a tool to help game designers to verify logical correctness of a game storyline, before starting the actual development by programmers. A domain specific language based on JavaScript Object Notation (JSON) has been here suggested by the author for storyline description. The Simple Promela Interpreter (SPIN) model checker has been  used for verification. As a case study, Fallout 3 game's storyline was considered and a logical error was found using the proposed methodology. Nevertheless, a verification approach is still required during implementation of a correct storyline by programmers insuring correctness of the code.

Computer games design is connected with simulation of the robotic systems. Thus it is important to take into consideration experience received from verification of robotic systems. One of the first attempts to apply formal verification to robotics systems was made in work \cite{SIM98}. The goal of authors was to develop a verification tool that can be used by people who are not familiar with formal verification methods to add reliability to design of robotic systems. The authors have proposed extension of the C++ programming language called TDL (Task Description language) to describe robotic system. The TDL program than can be translated to the model checker language to insure satisfiability of desired properties.

Robotic system can consist of interacting intelligent components or agents. Such systems are called multi-agent. Gaia methodology was introduced in \cite{WOO00} in order to provide an instrument for building a model of the multi-agent system from its specification. The model is based on role notion with attached responsibilities, permissions and activities. Responsibilities are introduced in order to specify goals of each role reaching which guarantees liveness (something good eventually happen) or safety (something bad never happens) properties of the multi-agent system. Target model gives high level understanding of the system and can help to insure logical correctness but does not impact significantly on reliability of the code.

Model Checking approach can be also a valuable tool for formal simulation and verification in swarms and swarm robotics, where multi-agent dynamic system behavior is frequently described by directed and undirected graphs and their geometric representations in a particular motion space \cite{Gazi2007}. 
These graphs $\mathcal{G}$ typically take into account the communication, information flow and interaction scenarios between agents $A$ in a swarm $\mathcal{S}$ by a weighted graph representation $\mathcal{G_S=(V_S,E_S,D_S)}$, where $\mathcal{V_S}$ is a vertex (node) set of agents $A_i$ with $i\in\mathcal{V_S}$, $\mathcal{E_S}$ is an edge set, which describes an information link $(i,j)\in\mathcal{E_S}$ between a pair $(A_i,A_j)$ of agents, and a weighting set $\mathcal{D_S}$ defines the desired weights $d_{i,j}\in\mathcal{D_S}$ that represent a control goal or a critical parameter (e.g., a distance between $A_i,A_j$). The structure and properties of $\mathcal{G_S}$ strictly depend on the hardware and software features of the swarm agents.

For formal verification of multi agent system models several well known model checkers can be applied: NuSMV \cite{CIM15}, PRISM \cite{KNP01}, MCMAS \cite{LOM17} and SPIN \cite{HOL97}. Interestingly to note that there is an extension of the PRISM model checker called PRISM-games \cite{KPW16} that can be helpful in finding optimal strategies for stochastic game models. However to apply model checker the model must be defined and there is no other way rather than explicitly providing all the formulas for description of the automata's states and transitions. For complex multi agent systems such straightforward approach can met the following difficulties:
\begin{itemize}
\item actions of actors can be too complex to provide symbolic representation for them by hand;\\
\item multi agent systems are usually parametrized and there is a need for fast regeneration of the model to adapt to new parameters; \\
\item an approach is needed for model reduction to address state explosion problem; \\
\end{itemize}

\section{Game Model} \label{sec:gmodel}

In this section, we will present the definitions necessary to construct a model of the game. We also give the definition of the evolution of the game.

Our definition of the multiplayer computer game is similar to definition of multi agent system \cite{RAI04} however it differs in several ways that are important for the model construction approach introduced in the next sections:
\begin{itemize}
\item Parametrization of the game is taken into account; \\
\item Actors can synchronize on actions (an actor can force another one to perform the same action); \\
\item Interaction between agents is taken into account through the collision operator; \\
\end{itemize}


\subsection*{Game Definition}
A multiplayer game is a tuple $\mathcal{G} = (\mathcal{A}, \mathcal{V}, \mathcal{V}_{init}, c, \mathcal{OP}, \mathcal{EV})$, where $\mathcal{A}$ is a set of actors; $\mathcal{V}$ is a set of vectors of attributes; $\mathcal{V}_{init}$ is the set of possible initial vectors of attributes; $c$ is a parameters tuple; $\mathcal{OP}$ is function that maps set of actions to each actor; and $\mathcal{EV}$ is the evolution operator that defines rules of changes of the game.

The mathematical modeling of a game is based on the concepts of \textit{actor}, \textit{attribute}, \textit{parameter} and \textit{action}, that will be described in detail in the following.

\subsection*{Actors}
Actors are all active objects which can influence a game process. They can be heroes representing players or landscape objects (for example, trees or walls) if they can perform actions in order to change something during the game. The set of all actors is denoted as $\mathcal{A}$. A concrete actor is denoted as $a\in\mathcal{A}$.

\subsection*{Attributes}
An attribute is an integer variable that might change its value during the game. For example, some actors have positions that can be changed during the game. To refer to a particular state of the game a tuple of concrete values of all attributes is considered. The set of all such possible tuples is denoted as 
\begin{equation*}
\mathcal{V} = \mathcal{V}_1\times\mathcal{V}_2\times\dots\times\mathcal{V}_m,
\quad\mathcal{V}_i\subset\mathbb{Z},\quad|\mathcal{V}_i| < \infty,\quad i=\overline{1,m}
\end{equation*}
A concrete vector of attributes is denoted as $v\in\mathcal{V}$. To refer to a particular attribute, its subscript index is used. For example, $v_1$ corresponds to the value of the first attribute of the vector $v = (v_1, v_2,\dots,v_m)$.
The game can be started from one of the initial vectors of attributes $v_{init}\in\mathcal{V}_{init}\subseteq{\mathcal{V}}$.

\subsection*{Parameters}

A parameter is a constant designated before the game starts and cannot be changed in future. Parameters do not change the nature of the game's actors and set of their actions, but change how these actions are performed. For example, size of the world or constant velocity of some actors. The tuple of parameters is denoted as $c = (c_1, c_2,\dots,c_k)$ and to refer to a particular parameter subscript index is used.

\subsection*{Definition (Projection)}
Projection of the set defined as the Cartesian product $X = X_1 \times X_2\times\dots\times X_n$ is a set $X^{\prime} = X_{i_1}\times X_{i_2}\times \dots \times X_{i_k}$ such that $k > 0$, $i_1 < i_2 < \dots < i_k$ and $i_s\in \big\{1,2,\dots,n\big\}, s = \overline{1,k}$. For simplicity the following denotation is used $Pr_{i_{1}i_{2}\dots i_{k}}(X) = X^{\prime}$.

Similarly, projection of the tuple $x = (x_1, x_2,\dots, x_k)$ is a tuple \\
$x^{\prime} = (x_{i_1},x_{i_2},\dots,x_{i_k})$ such that $k > 0$, $i_1 < i_2 < \dots < i_k$ and $i_s\in \big\{1,2,\dots,n\big\}, s = \overline{1,k}$. Projection of the tuple $x$ is denoted as $Pr_{i_{1}i_{2}\dots i_{k}}(x) = x^{\prime}$.

Let us denote $Ind(Pr_{i_{1}i_{2}\dots i_{k}}(X)) = \big\{i_{1},i_{2},\dots, i_{k}\big\}$.

\subsection*{Actions}

Let us consider the set of partial functions that depends on parameters tuple
$\mathcal{FP}_c = \big\{f_c: \mathcal{V}^{\prime}\rightarrow Pr_{i_{1}i_{2},\dots i_{k}}(\mathcal{V}),\quad k = \overline{1,m},\quad \mathcal{V^{\prime}\subseteq{V}} \big\}$. For a function $f\in\mathcal{FP}_c$, $D(f)$ and $R(F)$ denote the domain and range of $f$, respectively. Each function from the set $\mathcal{FP}_c$ is called action. As an example of such function can be a move action that changes the position of an actor.

Actions assigning operator is $\mathcal{OP}: \mathcal{A} \rightarrow 2^{\mathcal{FP}_{c}}$
such that:
\begin{enumerate}
\item for any actor $a\in\mathcal{A}: \bigcup\limits_{f\in\mathcal{OP}(a)}D(f) = \mathcal{V}$ 
\item for any actors $a^{\prime},a^{\prime\prime}\in\mathcal{A}$ and any functions $f^{\prime}\in\mathcal{OP}(a^{\prime}),f^{\prime\prime}\in\mathcal{OP}(a^{\prime\prime})$ one of the following must be true:
\begin{itemize}
\item $f^{\prime} = f^{\prime\prime}$
\item $D(f^{\prime}) \cap D(f^{\prime\prime}) = \emptyset$
\item $Ind(R(f^{\prime})) \cap Ind(R(f^{\prime\prime})) = \emptyset$
\end{itemize}
\end{enumerate}

The first part of the definition states that each actor must have at least one performable action for each possible attribute vector and the second part states that actors can share actions, but different actions performable at the same time cannot change the same attributes.

\subsection*{Game evolution}
We finally need to define the game evolution, how the game can progress starting from an initial vector of attributes belonging to $\mathcal{V}_{init}$.


Lets assume $\mathcal{A}$ is the set of actors in the game and lets consider the set $\mathcal{CP}_c = \big\{g_c:V^{\prime}\times \mathcal{FP}_{c}^{n} \rightarrow Pr_{i_{1}i_{2},\dots i_{k}}(\mathcal{V}),\quad k = \overline{1,m},\quad \mathcal{V^{\prime}\subseteq{V}} \big\}$, $g\in\mathcal{CP}_c$ is called a \textbf{collision} operator.

The operator $\mathcal{EV}:\mathcal{V}\rightarrow\mathcal{V}$ is called \textbf{evolution} if for any $v\in\mathcal{V}$ there is 
\begin{itemize}
\item a set of actions\\ 
$Act=\big\{f_{1},f_{2},\dots,f_{n},\ f_{i}\in \mathcal{OP}(a_{i}),\ a_i \in A,\ i=\overline{1,n}\big\}$, 
\item a set of 
collision operators \\
$Col=\big\{g_1,g_2,\dots,g_k,\ g_i\in\mathcal{CP}_c,\ i = \overline{1,k}\big\}$, and
\item partitions $I_{1},I_{2},\dots,I_{r}$ of the set $I = \big\{1,2,\dots,m\big\}$
\end{itemize}
such that
\begin{enumerate}
\item $\forall{f}\in Act\quad \exists{j} \in \big\{1,2,\dots,r\big\}$ such that $I_j = Ind(R(f))$
\item $\forall{j}\in \big\{1,2,\dots,r\big\}$ one of the following is true:
\begin{itemize}
\item $I_j = Ind(R(f)),\quad f\in Act$ 
\item $I_j = Ind(R(g)),\quad g\in Col$ 
\end{itemize}
\end{enumerate}

In other words, after each period of time, every player must perform an action. Actions are performed simultaneously and, in result, some of the attributes are attempted to be changed. The attributes which are not updated by the action functions are updated by the collision operators to fix collisions which could occur in the game after the actors' actions.

\subsection*{Kripke Structure}
A Kripke structure \cite{KRI63} is usually considered as a convenient way of the software and hardware model representation to further application of the verification procedure by a model checker tool. Thus in order to apply Model Checking for the verification of multiplayer games, it is necessary to formally define a Kripke structure from $\mathcal{G} = (\mathcal{A}, \mathcal{V}, \mathcal{V}_{init}, c, \mathcal{OP}, \mathcal{EV})$.

A \textit{Kripke structure} corresponding to a multiplayer game $\mathcal{G}$ is a tuple $K_{\mathcal{G}} = (S, S_0, \delta, AP, L)$ where $S$ is a finite set of states; $S = \mathcal{V}$; $S_0$ -  a set of initial states, $S_0 = \mathcal{V}_{init}$; $\delta: S \rightarrow S$ is a transition operator, $\delta = \mathcal{EV}$; $AP$ is a set of propositions; $L: S \rightarrow 2^{AP}$ is a label function that assigns the set of propositions to each state. Label function is defined by representing each proposition as first order predicates over the attributes of $\mathcal{G}$.

The objective of the Model Checking is ensuring that the model represented by the Kripke structure $K_{\mathcal{G}}$ satisfies a property defined in terms of Temporal Logic \cite{CLA86}.

\section{The Verification Method}\label{sec:vm}
In this section we describe the general method to  integrate Model Checking into game development process. We also show the case study of the \textit{Penguin Clash} game.

\subsection{Application to multiplayer game development}
\begin{figure}[ht!]
\includegraphics[width=\linewidth]{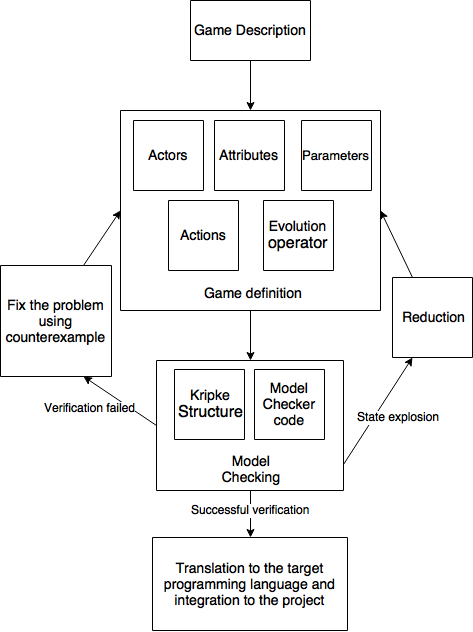}
\caption{Integration of the Model Checking into multiplayer game development process}
\label{fig:sgmet}
\end{figure}

Figure \ref{fig:sgmet} summarizes the proposed method of application of Model Checking in multiplayer game development. There are a few key steps to perform in order to apply the proposed method:

\begin{enumerate}
\item From the game description construct a game definition as discussed in Section \ref{sec:gmodel}
\item The game definition allows the construction of a Kripke structure for the specific game as discussed in Section \ref{sec:gmodel}
\item The verification can fail or be impossible because of \textit{state explosion} 
	\begin{enumerate}
	\item In case of failure the model checker will provide a counterexample to be used in the debugging phase
    \item In case of \textit{state explosion} a \textit{reduction technique} should be applied, see Section \ref{sec:cstudy}.
	\end{enumerate}
    \item Translation to a target programming language and integration of the verified NFA into a project.
\end{enumerate}

The integration of the verified NFA of the game into the project, which mainly consists in the translation of the model checker code to target the game development programming language, will not be discussed here, and it is left as future work. The game model construction and reduction procedure that helps to cope with the state explosion problem are instead discussed in detail in the next section.

\subsection{Case Study: Penguin Clash}\label{sec:cstudy}

The goal of the case study is to apply the proposed method for building a verified model for a real multiplayer game: \textit{Penguin Clash}. NuSMV model checker tool \cite{CIM15} is used to verify different properties of the result model.

\subsection*{Description}

\begin{figure}[ht!]
\includegraphics[width=\linewidth]{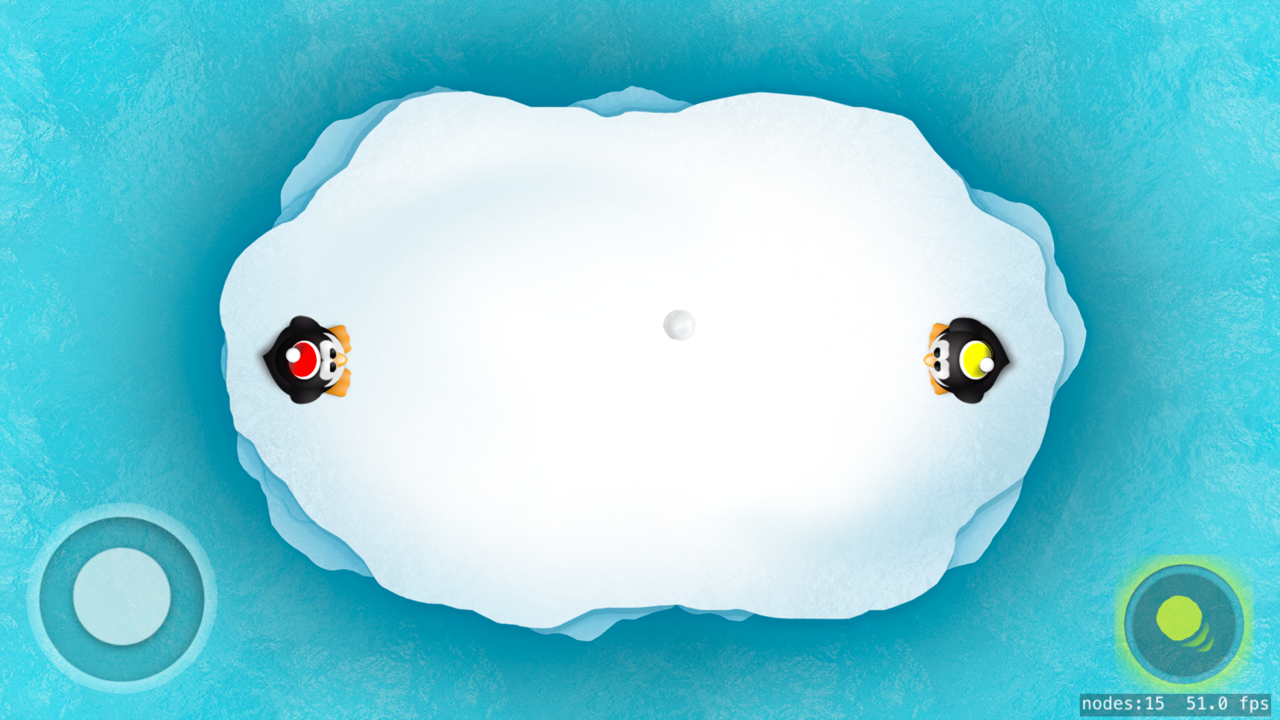}
\caption{Penguin Clash game design}
\end{figure}

\textit{Penguin Clash} is a multiplayer computer game. Each player controls a penguin that is  located on a ellipsis-shaped island surrounded by the ocean. Penguins can \textit{move} at a constant speed, \textit{throw} snowballs and perform \textit{flash} action to push the opponent. If a snowball thrown by one penguin hits the other one then the snowball is destroyed and the target penguin is stunned for a while. When a penguin is stunned it can do nothing. If a thrown snowball reaches the ocean then it is destroyed. Snowballs have a cooldown time: a penguin must wait for a while to throw a snowball. The cooldown time is larger than the maximum time between throw and destroy events. This means that there cannot be a situation when there are two snowballs thrown by a particular penguin. Also each penguin can accelerate to push the opponent. This is called a \textit{flash} action. If a penguin during flash action \textit{collides} with the second one the first penguin stops and the second one is \textit{pushed}. Velocity of the pushed penguin depends on the collision angle and decrements to simulate sliding on the ice. When a penguin is pushed it cannot perform flash or move actions, but it can throw the snowball. If penguins collide during move action or when both of them perform flash action than they just move in opposite directions with the same speed. If one of the penguins falls into the ocean it \textit{dies} and the game finishes.

\subsection*{Model}

Let us create the game model $\mathcal{G} = (\mathcal{A}, \mathcal{V}, \mathcal{V}_{init}, c, \mathcal{OP}, \mathcal{EV})$ from the description.

There are four actors: two penguins and two snowballs. Lets denote penguins as $pg^{(i)}$ and snowballs as $sb^{(i)}$ where $i = \overline{1,2}$. Then the set of actors is defined as follows $\mathcal{A}=\big\{pg^{(1)},sb^{(1)},pg^{(2)},sb^{(2)}\big\}$.

The set of attribute vectors $\mathcal{V}$ is defined using projections to the attributes belonging to penguin and snowball correspondingly. Tables \ref{tab:pgattr} and \ref{tab:sbattr} provide such description of the attributes.

\begin{table*}
\centering
\caption{Attributes of a penguin}
\smallskip
\begin{tabular}{| m{2cm} | m{5cm} | m{5cm} |}
  \hline
  \label{tab:pgattr}
  \textbf{Attribute} & \textbf{Domain} & \textbf{Description} \\ \hline
  $pg_{x}^{(i)}$ & $ 0 \le pg_{x}^{(i)} \le w_{xmax}$ &
  Penguin's x coordinate in 2D world \\ \hline
  $pg_{y}^{(i)}$ & $ 0 \le pg_{y}^{(i)} \le w_{ymax}$ &
  Penguin's y coordinate in 2D world \\ \hline
  $pg_{d}^{(i)}$ & $ 0 \le pg_{d}^{(i)} < 360$ &
  Direction of the last penguin's action \\ \hline
  $pg_{tstun}^{(i)}$ & $0 \le pg_{tstun}^{(i)} \le pg_{tstunmax} $ &
  Time during which the penguin remains stunned \\ \hline
  $pg_{iflash}^{(i)}$ & $0 \le pg_{iflash}^{(i)} \le pg_{iflashmax} $ &
  Countdown index determines current penguin's velocity during flash action \\ \hline
  $pg_{tflash}^{(i)}$ & $0 \le pg_{tflash}^{(i)} \le pg_{tflashmax} $ &
  Time remained during which penguin can not perform the flash action \\ \hline
  $pg_{ipushed}^{(i)}$ & $0 \le pg_{ipushed}^{(i)} \le pg_{ipushedmax} $ &
  Countdown index determines current penguin's velocity after it was pushed by another penguin \\ \hline
  $pg_{vpushed}^{(i)}$ & $0 \le pg_{vpushed}^{(i)} \le pg_{vpushedmax} $ &
  Initial velocity of the penguin when it was pushed by another penguin \\ \hline
  $pg_{dead}^{(i)}$ & $pg_{dead}^{(i)} \in \big\{0,1\big\} $ &
  Flag that states whether penguin is dead\\ \hline
  $pg_{pg^{(j)}}^{(i)}$ & $pg_{pg^{(j)}}^{(i)} \in \big\{0,1\big\} $ &
  Flag that states whether a penguin is colliding with another penguin\\ \hline
  $pg_{sb^{(j)}}^{(i)}$ & $pg_{sb^{(j)}}^{(i)} \in \big\{0,1\big\} $ &
  Flag that states whether a penguin is colliding with the snowball thrown by another penguin\\ \hline
  $pg_{tsnowball}^{(i)}$ & $0 \le pg_{tsnowball}^{(i)} \le pg_{tsnowballmax} $ &
  Time during which the penguin can not throw the snowball \\ \hline
\end{tabular}
\end{table*}

\begin{table*}
\centering
\caption{Attributes of a snowball}
\smallskip
\begin{tabular}{| m{2cm} | m{5cm} | m{5cm} |}
  \hline
  \label{tab:sbattr}
  \textbf{Attribute} & \textbf{Domain} & \textbf{Description} \\ \hline
  $sb_{x}^{(i)}$ & $ 0 \le sb_{x}^{(i)} \le w_{xmax}$ &
  Snowball's x coordinate in 2D world \\ \hline
  $sb_{y}^{(i)}$ & $ 0 \le sb_{y}^{(i)} \le w_{ymax}$ &
  Snowball's y coordinate in 2D world \\ \hline
  $sb_{d}^{(i)}$ & $ 0 \le sb_{d}^{(i)} < 360$ &
  Snowball's flying direction \\ \hline
  $sb_{flying}^{(i)}$ & $ sb_{flying}^{(i)} \in \big\{0,1\big\}$ &
  Flag that states whether snowball was thrown and haven't been destroyed yet.  \\ \hline
\end{tabular}
\end{table*}

The set of initial attribute vectors $\mathcal{V}_{init}$ consists of one element describing the state where penguins are located on the opposite sides of the island and no snowballs have been thrown yet.

The parameter vector $c$ with description of the components is provided in Table \ref{tab:gameparam}.

\begin{table*}
\centering
\caption{Actions of a snowball}
\smallskip
\begin{tabular}{| m{2cm} | m{10.5cm} |}
\hline
\label{tab:sbactions}
\textbf{Action} & \textbf{Description} \\ \hline
Deact & Action changes nothing keeping the fly flag false \\ \hline
Fly & Action changes the coordinates of a snowball when the fly flag is true and there is no collision with the opponent penguin or island border \\ \hline
Collide & Action is aimed to destroy a snowball changing the fly flag to false after collision with the opponent penguin \\ \hline
Throw & The same action as in Table \ref{tab:pgactions} \\ \hline
ThrowWhilePushed & The same action as in Table \ref{tab:pgactions} \\
\hline
\end{tabular}
\end{table*}

\begin{table*}
\centering
\caption{Actions of a penguin}
\smallskip
\begin{tabular}{| m{4cm} | m{8.5cm} |}
\hline
\label{tab:pgactions}
\textbf{Action} & \textbf{Description} \\ \hline
Move & Changes the coordinates of a penguin based on $pg_{velocity}$ \\ \hline
Throw & A penguin which is not pushed can perform this action by placing the snowball in the world and changing it's flying flag to true \\ \hline
ThrowWhilePushed & Meaning of the action is the same as of previous one but it can be performed when a penguin is pushed \\ \hline
Push & Initiates flashing movement of a penguin to push another one \\ \hline
Stay & Action which keeps the coordinates of a penguin unchanged \\ \hline
Flash & Changes the coordinates of a penguin during the flash action based on non-zero flashing index attribute \\ \hline
Pushed & Changes the coordinate of a penguin based on the non-zero pushed index attribute\\ \hline
CollidePenguin & Initializes the push index and the direction attributes after collision between penguins\\ \hline
CollideSnowball & Initializes the stun timer attribute of a penguin after collision with a snowball\\ \hline
CollideSnowballPenguin & Action is performed when collision with the opponent and it's snowball is detected at the same time. Thus the action combines logic of the previous two actions\\ \hline
Dead & This is the only action that can be performed when the dead attribute of a penguin is true and does not change any attribute\\ \hline
\end{tabular}
\end{table*}

\begin{table*}
\centering
\caption{Parameters of the game}
\smallskip
\begin{tabular}{| m{2cm} | m{10.5cm} |}
  \hline
  \label{tab:gameparam}
  \textbf{Parameter} & \textbf{Description} \\ \hline
  $w_{xmax}$ & Width of the 2D world \\ \hline
  $w_{ymax}$ & Height of the 2D world \\ \hline
  $isd_{xcenter}$ & X coordinate of the center of the ellipsis island in the 2D world \\ \hline
  $isd_{ycenter}$ & Y coordinate of the center of the ellipsis island in the 2D world \\ \hline
  $isd_{bradius}$ & Big radius of the ellipsis island in the 2D world \\ \hline
  $isd_{sradius}$ & Small radius of the ellipsis island in the 2D world \\ \hline
  $isd_{friction}$ & Friction coefficient of the ellipsis island in the 2D world \\ \hline
  $pg_{radius}$ & Radius of the body of the penguin  \\ \hline
  $pg_{velocity}$ & Number of points penguin can move per unit of time  \\ \hline
  $pg_{tstunmax}$ & Time during which penguin can perform no action if it has been hit by a snowball  \\ \hline
  $pg_{iflashmax}$ & Maximum flash index value \\ \hline
  $pg_{vflash}$ & Initial velocity of the penguin after performing flash action \\ \hline
  $pg_{tflashmax}$ & Time during which a penguin can not perform flash action after previous one \\ \hline
  $pg_{tsnowballmax}$ & Time during which a penguin can not throw a snowball after throwing previous one \\ \hline
  $sb_{radius}$ & Radius of a snowball \\ \hline
  $sb_{vfly}$ & Number of points a snowball can fly during the unit of time \\ \hline
\end{tabular}
\end{table*}

Finally, the set of actions $\mathcal{OP}$ for each actor are defined in Table \ref{tab:sbactions} and Table \ref{tab:pgactions}.

It can be noticed that a penguin and the snowball corresponding to it shares \textit{Throw} and \textit{ThrowWhilePushed} actions. Each of the actions can be performed at the same time by the actors and this situation is allowed by the definition of the $\mathcal{OP}$ operator.

There are attributes that cannot be modified by the actions: $pg_{pg^{(j)}}^{(i)}$, $pg_{sb^{(j)}}^{(i)}$, $pg_{dead}^{(i)}$. The attributes are updated by the collision operators to insist on performing push actions by the penguins when they collide, to stun a penguin after collision with a snowball and to kill a penguin by detecting intersection with the island's border. Also there is a collision operator which detects snowball intersection of the island's border and destroys it by setting it's fly flag to false.

NuSMV is used for this case study to verify properties of the game model. It is based on the symbolic model checking approach \cite{CLA90}. In other words, the set of all states, the set of initial states and the transition function of the Kripke structure are transformed to be represented using boolean functions. Properties to verify for this case study are provided to the model checker as formulas in CTL Temporal Logic \cite{CLA81}. NuSMV model checker uses Binary Decision Diagrams (BDD) \cite{BRY86} to store and perform operations on boolean functions that is a key ingredient of the symbolic model checking approach.

Representation of the actions by boolean functions is not given in this article due to space constraints. The full source code of the case study is available on Github \cite{SG17}.

\subsection*{Verification results}

The first property which was attempted to be verified for the previously described model can be represented in CTL as follows:
\begin{equation*}
AG\quad EF\quad pg_{dead}^{(1)}\quad \vee\quad pg_{dead}^{(2)}
\end{equation*}

The meaning of the formula is that for any future state of the game ($AG$) there is always an opportunity ($EF$) which leads to the state where at least one of the penguins will be dead: leads to the game completion.

\begin{figure}[ht!]
\includegraphics[width=\linewidth]{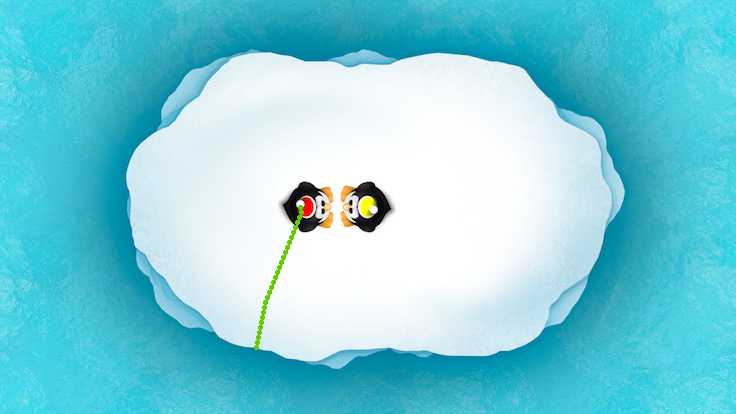}  
\caption{Verification results with counterexample of a movement behavior (green line)}
\label{fig:pgvermv}
\end{figure}

Originally the case study was inspired by \cite{BUT91}, which discusses successful verification applied to the model with $10^{120}$ states. The model for \textit{Penguin Clash} game has $\approx 10^{72}$ states. Unfortunately, the size of the set of states was too big for the model checker to process. 

This problem can be met very often in the model checking world and is known as state explosion problem. To address the issue, reduction procedure \cite{KEI13} is usually applied to the model. In result, smaller model is received which is equivalent to the original model in terms of the property to be verified. This means that property correctness result is the same for both models.

In case of the game model $\mathcal{G} = (\mathcal{A}, \mathcal{V}, \mathcal{V}_{init}, c, \mathcal{OP}, \mathcal{EV})$ reduction procedure can be defined by projection operation applied to the set $\mathcal{V}$ and removal operation applied to the set of actions of each actor.

\begin{figure}[ht!]
\includegraphics[width=\linewidth]{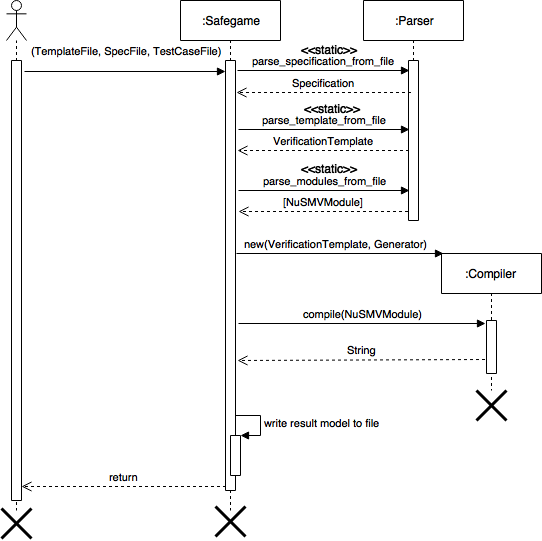}
\caption{Dynamic view of the Safegame tool}
\label{fig:genseq}
\end{figure}

During the case study the \texttt{Safegame} tool was developed. High level view on its functioning is provided on Fig. \ref{fig:genseq}.

 Main purpose of the tool is to provide a flexible instrument to create reduced models from the original one, allow implementation of custom functions in Generator component which automatically replace parameters of the model with concrete values and insert symbolic expressions of complex actions into the model. 

Based on formulated requirements input of the tool is represented by three parts: path to parametrized game model (template) file, path to game specification file and path to file with test cases.

Template file contains code in NuSMV language representing the game model, but explicit values of actions and parameters are substituted with tags. Tag is a valid identifier (can contain ascii letters or \_) surrounded with @ symbol. For example, @moved@.

Specification file contains text in YAML (YAML Ain't Markup Language) \cite{YA09} format defining arguments needed to generate values for parameters from the template. Also specification file can contain a list of (key, value) pairs to provide default values for parameters and actions. Using the fact that actions are represented by first order predicate logic formulas we can provide constant default expression (true or false) for those actions which we know don't influence correctness of the property but from the other hand contribute to the number of states of the model. The strategy helps to  struggle with the state explosion problem.

Last ingredient to generate final game model is a test case. Test case is represented as NuSMV module containing game's initial configuration, transition block and definition of the LTL or CTL properties to verify.

Parser component is responsible for parsing of input files to create internal object. The task of Generator component is to create data from the specification. For example, set of all possible position offsets  of the penguin to move is generated based on it's velocity. Finally, Compiler class object is created to generate final game model code for NuSMV model checker from parsed specification, template and test case using generator. Tags  are replaced either with default expressions from the Specification  or with generated ones. Interface of the Generator component consists of methods with the same name as action tags. This approach allows to reach independence between Compiler and Generator through well known reflection mechanism. It can be noted that Generator is the only component that depends on the concrete game and other ones can be reused.

For tool demonstration it was assumed that first penguin can perform Move, Stay, Collide and Dead actions and the second one can perform only Stay and Collide actions. Figure \ref{fig:pgvermv} shows the initial state of the game. For this reduced model the following properties were verified:
\begin{enumerate}
\item $EF\quad pg_{dead}^{(1)}$
\item $AG\quad \neg pg_{dead}^{(2)}$
\item $AG\quad EF\quad pg_{pg^{(2)}}^{(1)}\quad \wedge\quad pg_{pg^{(2)}}^{(1)}$
\item $AG\quad \neg pg_{dead}^{(1)}\quad \Rightarrow \quad EF\quad (pg_{pg^{(2)}}^{(1)}\quad \wedge \quad pg_{pg^{(2)}}^{(1)})$
\end{enumerate}



The first property checks that the first penguin can reach the ocean and die. The second one verifies that the second penguin cannot die. Properties 3 and 4 check collision behavior: property 3 checks that from any future state there is an opportunity to reach the state where penguins collide each other; property 4 verifies the same but with precondition that the first penguin must not be dead. The model checker successfully verified properties 1, 2 and 4. The model checker was not able to verify property 3, however it provided a counterexample which is depicted on Figure \ref{fig:pgvermv}. The green line shows the changes of the position and the direction of the first penguin until it reaches the state where its dead flag is true. Due to the fact that first penguin cannot move because it is dead and second one cannot move by definition, the game cannot reach a collision state. Thus the property does not hold.

Model checking process was executed on the computer: 2,3 GHz Intel Core i5 8GB RAM. It took about 2,5 hours to complete verification. Number of states of the target NFA was $\approx 10^{9}$.

\section{Conclusion}
\label{concl}


Game design support is an emerging and active research area, still lacking formal tools that have been already widely accepted and adopted in other scientific field and industries. In this paper, we attacked this problem and we contributed with a novel approach to formally verify computer games. We propose a method of model construction that starts from a computer game description and  utilizes Model Checking. The method was applied to a case study: the multiplayer game \textit{Penguin Clash}. A solution for the state explosion problem (using \texttt{Safegame}, a tool developed for this purpose) was offered by means of an example of verification of properties related to movement and collision. 

Although the verification process executed during the case study was not fast, the considered technique will be further studied and developed in order to be applied in a synergistic way with more popular methods used in TDD, and therefore providing more reliable testing procedure. The \texttt{Safegame} tool can be taken as a basis for further development.

Another direction for future work can be the probabilistic extension of the defined game model to allow verification of quantitative properties. For example, actor's action could be considered as distribution of a random variable. This extension could help approaching problems of Game Theory. For example, strategy retrieval corresponding to an equilibrium state, as studied in \cite{KWI16}.

Robotic multi-agent systems, consisting of interacting intelligent components, may also be naturally modeled following a similar approach. Swarm robotics appear to be the most immediate application.

\bibliography{main}
\bibliographystyle{abbrv}

\end{document}